\documentclass[letterpaper,12pt]{article}
\usepackage[margin=1in]{geometry}
\usepackage{verbatim}
\usepackage{amsmath}
\usepackage{amssymb}
\usepackage{graphicx}
\usepackage{amsthm}
\usepackage{subfig}
\usepackage[round,authoryear]{natbib}

\def\epsilon{\varepsilon}

\def\0s{{\bf 0}}

\newtheorem{theorem}{Theorem}[section]

\newtheorem{observe}[theorem]{Observation}
\newtheorem{remark1}[theorem]{Remark}

\newenvironment{remark}{\begin{remark1} \rm}{\end{remark1}}

\title{Significance testing without truth}
\author{William Perkins, Mark Tygert, and Rachel Ward}

\begin{document}

\maketitle

\begin{abstract}
A popular approach to significance testing proposes to decide whether
the given hypothesized statistical model is likely to be true (or false).
Statistical decision theory provides a basis for this approach
by requiring every significance test to make a decision about the truth
of the hypothesis/model under consideration.
Unfortunately, many interesting and useful models are obviously false
(that is, not exactly true) even before considering any data.
Fortunately, in practice a significance test need only gauge the consistency
(or inconsistency) of the observed data with the assumed hypothesis/model
--- without enquiring as to whether the assumption is likely to be true
(or false), or whether some alternative is likely to be true (or false).
In this practical formulation, a significance test rejects a hypothesis/model
only if the observed data is highly improbable when calculating
the probability while assuming the hypothesis being tested;
the significance test only gauges whether the observed data likely
invalidates the assumed hypothesis, and cannot decide that the assumption
--- however unmistakably false --- is likely to be false a priori,
without any data.
\end{abstract}

\begin{quotation}
{\it Essentially, all models are wrong, but some are useful.} --- G.~E.~P. Box
\end{quotation}

\tableofcontents

\section{Introduction}

As pointed out in the above quotation of G.~E.~P. Box,
many interesting models are false (that is, not exactly true),
yet are useful nonetheless.
Significance testing helps measure the usefulness of a model.
Testing the validity of using a model for virtually any purpose
requires knowing whether observed discrepancies are due to inaccuracies
or inadequacies in the model or (on the contrary) could be due
to chance arising from necessarily finite sample sizes.
Significance tests gauge whether the discrepancy between the model
and the observed data is larger than expected random fluctuations;
significance tests gauge the size of the unavoidable random fluctuations.

A traditional approach, along with its modern formulation
in statistical decision theory, tries to decide whether a hypothesized model
is likely to be true (or false).
However, in many practical circumstances, a significance test need only gauge
the consistency (or inconsistency) of the observed data
with the assumed hypothesis/model
--- without ever enquiring as to whether the assumption is likely to be true
(or false), or whether some alternative is likely to be true (or false).
In this practical formulation, a significance test rejects a hypothesis/model
only if the observed data is highly improbable when calculating
the probability while assuming the hypothesis being tested.
Whether or not the assumption could be exactly true in reality is irrelevant.

An illustrative example may help clarify.
When testing the goodness of fit for the Poisson regression
where the distribution of $Y$ given $x$ is the Poisson distribution of mean
$\exp(\theta^{(0)}+\theta^{(1)}x+\theta^{(2)}x^2+\theta^{(3)}x^3)$,
the conventional Neyman-Pearson null hypothesis is
\begin{multline}
\label{theirs}
H^{\rm NP}_0 : \hbox{there exist real numbers }
  \theta^{(0)}, \theta^{(1)}, \theta^{(2)}, \theta^{(3)} \hbox{ such that }
  y_1, y_2, \dots, y_n \hbox{ are independent}\\
  \hbox{ draws from the Poisson distributions with means }
  \mu_1, \mu_2, \dots, \mu_n, \hbox{ respectively,}
\end{multline}
where
\begin{equation}
\label{theirsaux}
\ln(\mu_k) = \theta^{(0)} + \theta^{(1)} x_k
           + \theta^{(2)} (x_k)^2 + \theta^{(3)} (x_k)^3
\end{equation}
for $k = 1$,~$2$, \dots, $n$,
and the observations $(x_1,y_1)$,~$(x_2,y_2)$, \dots, $(x_n,y_n)$
are ordered pairs of scalars (real numbers paired with nonnegative integers).
A related but perhaps simpler null hypothesis is
\begin{multline}
\label{ours}
H_0 : y_1, y_2, \dots, y_n
  \hbox{ are independent draws from the Poisson distributions}\\
  \hbox{with means }
  \hat\mu_1, \hat\mu_2, \dots, \hat\mu_n,\hbox{ respectively,}
\end{multline}
where
\begin{equation}
\label{oursaux}
\ln(\hat\mu_k) = \hat\theta^{(0)} + \hat\theta^{(1)} x_k
               + \hat\theta^{(2)} (x_k)^2 + \hat\theta^{(3)} (x_k)^3
\end{equation}
for $k = 1$,~$2$, \dots, $n$,
with $\hat\theta$ being a maximum-likelihood estimate.
Needless to say, even if the observed data really does arise
from Poisson distributions whose means are exponentials
of a cubic polynomial, the particular values
$\hat\theta^{(0)}$,~$\hat\theta^{(1)}$, $\hat\theta^{(2)}$,~$\hat\theta^{(3)}$
of the parameters of the fitted polynomial will almost surely not be
exactly equal to the true values.
Even though the estimated values of the parameters may not be exactly correct,
it still makes good sense to enquire as
to whether the fitted cubic polynomial is consistent
with the data up to random fluctuations inherent
in using a finite amount of observed data.

In fact, since subsequent use of the model
usually involves the particular fitted polynomial ---
whose specification includes the observed parameter estimates ---
analyzing the model including the estimated values of the parameters
makes more sense than trying to decide whether the data really
did come from Poisson distributions whose means
are exponentials of some unspecified cubic polynomial.
For instance, any plot of the fit (such as a plot of the means
of the Poisson distributions) must use the estimated values
of the parameters, and any statistical interpretation of the plot should also
depend explicitly on the estimates;
a significance test can gauge the consistency of the plotted fit
with the observed data, without ever asking whether the plotted fit
is the truth (it is almost surely not identical to the underlying reality)
and without making some decision about an abstract family of polynomials
which may or may not include both the plotted fit and the underlying reality.

A popular measure of divergence from the null hypothesis
is the log--likelihood-ratio
\begin{equation}
g^2 = 2\sum_{k=1}^n y_k \ln(y_k/\hat\mu_k).
\end{equation}
A P-value (see, for example, Section~\ref{Pvalues} below)
quantifies whether this divergence is larger than expected
from random fluctuations inherent in using only $n$ data points.
It is not obvious how to calculate an exact P-value for $H^{\rm NP}_0$
from~(\ref{theirs}) and~(\ref{theirsaux}),
which refers to cubic polynomials with undetermined coefficients.
In contrast, $H_0$ from~(\ref{ours}) and~(\ref{oursaux})
refers explicitly to the particular fitted value $\hat\theta$;
$H_0$ concerns the particular fit displayed in a plot,
and is natural for the statistical interpretation of such a plot.

Thus, when calculating significance,
the assumed model should include the particular values of any parameters
estimated from the observed data.
Such parameters are known as ``nuisance'' parameters.
As illustrated with $H_0$ from~(\ref{ours}) and~(\ref{oursaux}),
the assumed hypothesis will be ``simple'' in the Neyman-Pearson sense,
but will depend on the observed values of the parameters ---
that is, the hypothesis will be ``data-dependent'';
the hypothesis will be ``random.''
Including the particular values of the parameters estimated
from the observed data replaces the ``composite'' hypothesis
of the conventional Neyman-Pearson formulation
with a ``simple'' data-dependent hypothesis.
As discussed in Section~\ref{distribution} below,
fully conditional tests also incorporate the observed values of the parameters,
but make the extra assumption that all possible realizations of the experiment
--- observed or hypothetical --- generate the same observed values 
of the parameters.
The device of a ``simple data-dependent hypothesis'' such as $H_0$ incorporates
the observed values explicitly without the extra assumption.

For most purposes, a parameterized model is not really operational
--- that is, suitable for making precise predictions ---
until its specification is completed via the inclusion of estimates
for any nuisance parameters.
The results of the significance tests considered below
depend on the quality of both the models and the parameter estimators.
However, the results are relatively insensitive
to the particular observed realizations
of the parameter estimators (that is, to the parameter estimates) unless
specifically designed to quantify the quality of the parameter estimates.
To quantify the quality of the parameter estimates,
we recommend testing separately the goodness of fit of the parameter estimates,
using confidence intervals, confidence distributions, parametric bootstrapping,
or significance tests within parametric models,
whose statistical power is focused against alternatives
within the parametric family constituting the model
(for further discussion of the latter, see Section~\ref{many} below).

The remainder of the present article has the following structure:
Section~\ref{Bayesvsfreq} very briefly discusses Bayesian-frequentist hybrids,
referring for details to the definitive work of~\cite{gelman}.
Section~\ref{Pvalues} defines P-values --- also known as
``attained significance levels'' ---
which quantify the consistency of the observed data with the assumed models.
Section~\ref{distribution} details several approaches to testing
the goodness of fit for distributional profile.
Section~\ref{many} discusses testing the goodness of fit
for various properties beyond just distributional profile.

\cite{cox} details many advantages of interpreting significance
as gauging the consistency of an assumption/hypothesis
with observed data,
rather than as making decisions about the actual truth of the assumption.
However, significance testing is meaningless without any observations,
unlike purely Bayesian methods, which can produce results without any data,
courtesy of the prior (the prior is the statistician's system
of a priori beliefs, accumulated from prior experience, law, morality,
religion, etc., without reference to the observed data).
Significance tests are deficient in this respect.
Those interested in what is to be considered true in reality
and in making decisions more generally should use Bayesian
and sequential (including multilevel) procedures.
Significance testing simply gauges the consistency of models
with observed data;
generally significance testing alone cannot handle the truth.

\section{Bayesian versus frequentist}
\label{Bayesvsfreq}

Traditionally, significance testing is frequentist.
However, there exist Bayesian-frequentist hybrids known as
``Bayesian P-values'';
\cite{gelman} sets forth a particularly appealing formulation.
Bayesian P-values test the consistency of the observed data
with the model {\it used together with a prior} for nuisance parameters.
In contrast, the P-values discussed in the present paper
test the consistency of the observed data
with the model {\it used together with a parameter estimator}.
In the Bayesian formulation, a P-value depends explicitly on the choice
of prior;
in the formulation of the present paper, a P-value depends explicitly
on the choice of parameter estimator.
Thus, when there are nuisance parameters,
the two types of P-values test slightly different hypotheses
and provide slightly different information;
each type is ideal for its own set-up.
Of course, if there are no nuisance parameters, then Bayesian P-values
and the P-values discussed below are the same.

\section{P-values}
\label{Pvalues}

A P-value for a hypothesis $H_0$ is a statistic such that,
if the P-value is very small, then we can be confident that the observed data
is inconsistent with assuming $H_0$.
The P-value associated with a measure of divergence from a hypothesis $H_0$
is the probability that $D \ge d$,
where $d$ is the divergence between the observed
and the expected (with the expectation following $H_0$ for the observations),
and $D$ is the divergence between the simulated
and the expected (with the expectation following $H_0$ for the simulations,
and with the simulations performed assuming $H_0$).
When taking the probability that $D \ge d$, we view $D$ as a random variable,
while viewing $d$ as fixed, not random.
For example, when testing the goodness of fit for the model
of i.i.d.\ draws from a probability distribution $p_0(\theta)$,
where $\theta$ is a nuisance parameter that must be estimated
from the data, that is, from observations $x_1$,~$x_2$, \dots, $x_n$,
we use the null hypothesis
\begin{equation}
H_0: x_1, x_2, \dots, x_n \hbox{ are i.i.d.\ draws from }
     p_0(\hat\theta), \hbox{where } \hat\theta = \hat\theta(x_1,x_2,\dots,x_n).
\end{equation}
The P-value for $H_0$ associated with a divergence $\delta$
is the probability that $D \ge d$, where
\begin{itemize}
\item $d = \delta(\hat{p},p_0(\hat\theta))$,
\item $\hat{p}$ is the empirical distribution of $x_1$,~$x_2$, \dots, $x_n$,
\item $\hat\theta$ is the parameter estimate
obtained from the observed draws $x_1$,~$x_2$, \dots, $x_n$,
\item $D = \delta(\hat{P},p_0(\hat\Theta))$,
\item $\hat{P}$ is the empirical distribution of i.i.d.\ draws
$X_1$,~$X_2$, \dots, $X_n$ from $p_0(\hat\theta)$, and
\item $\hat\Theta$ is the parameter estimate
obtained from the simulated draws $X_1$,~$X_2$, \dots, $X_n$.
\end{itemize}
If the P-value is very small, then we can be confident that the observed data
is inconsistent with assuming $H_0$.
Examples of divergences include $\chi^2$ (for categorical data)
and the maximum absolute difference between cumulative distribution functions
(for real-valued data).

\begin{remark}
\label{MonteCarlo}
To compute the P-value assessing the consistency
of the experimental data with assuming $H_0$,
we can use Monte-Carlo simulations
(very similar to those of~\cite{clauset-shalizi-newman}).
First, we estimate the parameter $\theta$
from the $n$ given experimental draws, obtaining $\hat\theta$,
and calculate the divergence between the empirical distribution
and $p_0(\hat\theta)$. We then run many simulations.
To conduct a single simulation, we perform the following three-step procedure:
\begin{enumerate}
\item we generate $n$ i.i.d.\ draws according
      to the model distribution $p_0(\hat\theta)$,
      where $\hat\theta$ is the estimate calculated
      from the experimental data,
\item we estimate the parameter $\theta$ from the data
      generated in Step~1, obtaining a new estimate $\tilde\theta$, and
\item we calculate the divergence between the empirical distribution
      of the data generated in Step~1 and $p_0(\tilde\theta)$,
      where $\tilde\theta$ is the estimate calculated in Step~2
      from the data generated in Step~1.
\end{enumerate}
After conducting many such simulations, we may estimate the P-value
for assuming $H_0$ as the fraction of the divergences calculated in Step~3
that are greater than or equal to the divergence calculated
from the empirical data.
The accuracy of the estimated P-value
is inversely proportional to the square root
of the number of simulations conducted;
for details, see Remark~\ref{error-bars} below.
\end{remark}

\begin{remark}
\label{error-bars}
The standard error of the estimate from Remark~\ref{MonteCarlo}
for an exact P-value $P$
is $\sqrt{P(1-P)/\ell}$, where $\ell$ is the number
of Monte-Carlo simulations conducted to produce the estimate.
Indeed, each simulation has probability $P$ of producing a divergence
that is greater than or equal to the divergence corresponding
to an exact P-value of $P$.
Since the simulations are all independent, the number of the $\ell$ simulations
that produce divergences greater than or equal to that corresponding
to P-value $P$ follows the binomial distribution
with $\ell$ trials and probability $P$ of success in each trial.
The standard deviation of the number of simulations whose divergences
are greater than or equal to that corresponding to P-value $P$ is therefore
$\sqrt{\ell P (1-P)}$, and so the standard deviation
of the {\it fraction} of the simulations producing such divergences
is $\sqrt{P (1-P)/\ell}$. Of course, the fraction itself
is the Monte-Carlo estimate of the exact P-value
(we use this estimate in place of the unknown $P$
when calculating the standard error $\sqrt{P (1-P)/\ell}$).
\end{remark}

\section{Goodness of fit for distributional profile}
\label{distribution}

Given observations $x_1$,~$x_2$, \dots, $x_n$,
we can test the goodness of fit for the model
of i.i.d.\ draws from a probability distribution $p_0(\theta)$,
where $\theta$ is a nuisance parameter, via the null hypothesis
\begin{multline}
H_0: x_1, x_2, \dots, x_n \hbox{ are i.i.d.\ draws from }
     p_0(\hat\theta)\\\hbox{for the particular observed value of }
     \hat\theta = \hat\theta(x_1,x_2,\dots,x_n).
\end{multline}

The Neyman-Pearson formulation considers instead the null hypothesis
\begin{equation}
H^{\rm NP}_0: \hbox{there exists a value of } \theta \hbox{ such that }
      x_1, x_2, \dots, x_n \hbox{ are i.i.d.\ draws from } p_0(\theta).
\end{equation}

The fully conditional null hypothesis is
\begin{multline}
H^{\rm FC}_0: x_1, x_2, \dots, x_n \hbox{ are i.i.d.\ draws from }
       p_0(\hat\theta)\\\hbox{ and }
       \hat\theta = \hat\theta(x_1,x_2,\dots,x_n)
       \hbox{ takes the same value in all possible realizations}.
\end{multline}
That is, whereas $H_0$ supposes that the particular observed realization
of the experiment happened to produce a parameter estimate $\hat\theta$
that is consistent with having drawn the data from $p_0(\hat\theta)$,
$H^{\rm FC}_0$ assumes that every possible realization of the experiment
--- observed or hypothetical --- produces exactly the same parameter estimate.
Few experimental apparatus constrain the parameter estimate to always take
the same (a priori unknown) value during repetitions of the experiment,
as $H^{\rm FC}_0$ assumes.
Assuming $H^{\rm FC}_0$ amounts to conditioning on a statistic
that is minimally sufficient for estimating $\theta$; computing the associated
P-values is not always trivial.
Furthermore, the assumption that $H^{\rm FC}_0$ is true seems to be more
extreme, a more substantial departure from $H^{\rm NP}_0$, than $H_0$.
Finally, testing the significance of assuming $H_0$ would seem
to be more apropos in practice for applications in which the experimental
design does not enforce that repeated experiments always yield the same value
for $p_0(\hat\theta)$. We cannot recommend the use of $H^{\rm FC}_0$
in general. Unfortunately, $H^{\rm NP}_0$ also presents problems\dots.

If the probability distributions are discrete,
there is no obvious means for defining an exact P-value
for $H^{\rm NP}_0$ when $H^{\rm NP}_0$ is false;
moreover, any P-value for $H^{\rm NP}_0$ when $H^{\rm NP}_0$ is true
would depend on the correct value of the parameter $\theta$,
and the observed data does not determine this value exactly.
The situation may be more favorable when measuring discrepancies
with divergences that are ``approximately ancillary'' with respect to $\theta$,
but quantifying ``approximately'' seems to be problematic
except in the limit of large numbers of draws.
(Some divergences are asymptotically ancillary in the limit of large numbers
of draws, but this is not especially helpful,
as any asymptotically consistent estimator $\hat\theta$ converges
to the correct value in the limit of large numbers of draws;
$\theta$ is almost surely known exactly
in the limit of large numbers of draws,
so there is no benefit to being independent of $\theta$ in that limit.)
Section~3 of~\cite{robins-wasserman} reviews these and related issues.

\begin{remark}
\cite{romano}, \cite{henze}, \cite{bickel-ritov-stoker}, and others
have shown that the P-values for $H_0$ converge in distribution
to the uniform distribution over $[0,1]$
in the limit of large numbers of draws, when $H^{\rm NP}_0$ is true.
In particular, \cite{romano} and~\cite{henze} prove this convergence
for a wide class of divergence measures.
\end{remark}

\begin{remark}
The surveys of~\cite{agresti1} and~\cite{agresti2}
discuss exact P-values for contingency-tables/cross-tabulations,
including criticism of fully conditional P-values.
\cite{gelman} provides further criticism of fully conditional P-values.
\cite{ward} numerically evaluates the different types of P-values
for an application in population genetics.
Section~4 of~\cite{bayarri-berger2} and the references it cites
discuss the menagerie of alternative P-values proposed recently.
\end{remark}

\section{Goodness of fit for various properties}
\label{many}

For comparative purposes,
we first review the null hypothesis of the previous section
for testing the goodness of fit for distributional profile, namely
\begin{equation}
H_0: x_1, x_2, \dots, x_n \hbox{ are i.i.d.\ draws from }
     p_0(\hat\theta), \hbox{where } \hat\theta = \hat\theta(x_1,x_2,\dots,x_n),
\end{equation}
with $\theta$ being the nuisance parameter.
The measure of discrepancy for $H_0$ is usually taken to be a divergence
between the empirical distribution $\hat{p}$ and the model $p_0(\hat\theta)$
(in the continuous case in one dimension,
a common characterization of the empirical distribution
is the empirical cumulative distribution function;
in the discrete case, a common characterization of the empirical distribution
is the empirical probability mass function, that is,
the set of empirical proportions).
One example for $p_0$ is the Zipf distribution over $m$ bins
with parameter $\theta$, a discrete distribution
with the probability mass function
\begin{equation}
p_0^{(j)}(\theta) = \frac{C_{\theta}}{j^\theta}
\end{equation}
for $j = 1$,~$2$,~$3$, \dots, $m$, where the normalization constant is
\begin{equation}
\label{normalization}
C_{\theta} = \frac{1}{\sum_{j=1}^m j^{-\theta}}
\end{equation}
and $\theta$ is a nonnegative real number.

When testing the goodness of fit for parameter estimates,
we use the null hypothesis
\begin{equation}
H'_0: x_1, x_2, \dots, x_n \hbox{ are i.i.d.\ draws from }
      p_0(\phi_0,\hat\theta),
      \hbox{where } \hat\theta = \hat\theta(x_1,x_2,\dots,x_n),
\end{equation}
with $\theta$ being the nuisance parameter and $\phi$ being the parameter
of interest (and with $\phi_0$ being the value of $\phi$ assumed
under the model). Please note that $H_0$ and $H'_0$ are actually equivalent,
via the identification $p_0(\theta) = p_0(\phi_0,\theta)$.
However, the measure of discrepancy for $H'_0$ is usually taken to be
a divergence between $\hat\phi$ and $\phi_0$ rather than the divergence
between $\hat{p}$ and $p_0(\hat\theta)$ that is more natural for $H_0$.
Also, if $\phi$ is scalar-valued,
then confidence intervals, confidence distributions,
and parametric bootstrap distributions are more informative
than a significance test. A significance test is appropriate
if $\phi$ is vector-valued.  
One example for $p_0$ is the sorted Zipf distribution over $m$ bins
with $\theta$ being the power in the power law
and with the maximum-likelihood estimate $\hat\phi$ being a permutation
that sorts the bins into rank-order, that is,
$p_0$ is the discrete distribution with the probability mass function
\begin{equation}
p_0^{(j)}(\phi,\theta) = \frac{C_{\theta}}{(\phi(j))^\theta}
\end{equation}
for $j = 1$,~$2$,~$3$, \dots, $m$, where the normalization constant
$C_{\theta}$ is defined in~(\ref{normalization})
with $\theta$ being a nonnegative real number,
and $\phi$ is a permutation of the numbers $1$,~$2$, \dots, $m$.
The choice for $\phi_0$ that is of widest interest in applications
is the identity permutation (that is, the ``rearrangement'' of the bins
that does not permute any bins: $\phi_0(j) = j$ for $j = 1$,~$2$, \dots, $m$).

When testing the goodness of fit for the standard Poisson regression
with the distribution of $Y$ given $x$ being the Poisson distribution of mean
$\exp\left(\theta^{(0)}+\sum_{j=1}^m \theta^{(j)} x^{(j)}\right)$,
we use the null hypothesis
\begin{multline}
H''_0 : y_1, y_2, \dots, y_n
  \hbox{ are independent draws from the Poisson distributions with means }
\\\exp\left(\hat\theta^{(0)}+\sum_{j=1}^m \hat\theta^{(j)} x^{(j)}_1\right),\;
  \exp\left(\hat\theta^{(0)}+\sum_{j=1}^m \hat\theta^{(j)} x^{(j)}_2\right),\;
  \dots,\;
  \exp\left(\hat\theta^{(0)}+\sum_{j=1}^m \hat\theta^{(j)} x^{(j)}_n\right),
\\\hbox{ respectively,}
\end{multline}
where $\theta$ is the nuisance parameter
and $\hat\theta$ is a maximum-likelihood estimate.
The measure of discrepancy for $H''_0$ is usually taken to be
the log--likelihood-ratio (also known as the deviance)
\begin{equation}
g^2 = 2\sum_{k=1}^n y_k \ln(y_k/\hat\mu_k),
\end{equation}
where $\hat\mu_k$ is the mean of the Poisson distribution associated with $y_k$
in $H''_0$, namely,
\begin{equation}
\hat\mu_k = \exp\left(\hat\theta^{(0)}
                     +\sum_{j=1}^m \hat\theta^{(j)} x^{(j)}_k\right).
\end{equation}
One example is the cubic polynomial
\begin{equation}
\ln(\mu_k) = \theta^{(0)} + \theta^{(1)} x_k + \theta^{(2)} (x_k)^2
           + \theta^{(3)} (x_k)^3
\end{equation}
for $k = 1$,~$2$, \dots, $n$, which comes from the choice $m = 3$ and
\begin{equation}
x^{(1)}_k = x_k; \quad x^{(2)}_k = (x_k)^2; \quad x^{(3)}_k = (x_k)^3
\end{equation}
for $k = 1$,~$2$, \dots, $n$,
given observations as ordered pairs of scalars
$(x_1,y_1)$,~$(x_2,y_2)$, \dots, $(x_n,y_n)$.
Of course, there are similar formulations for other generalized linear models,
such as those discussed by~\cite{mccullagh-nelder}.

\section*{Acknowledgements}
\addcontentsline{toc}{section}{\protect\numberline{}Acknowledgements}

We would like to thank Alex Barnett, Andrew Barron, G\'erard Ben Arous,
James Berger, Tony Cai, Sourav Chatterjee, Ronald Raphael Coifman,
Ingrid Daubechies, Jianqing Fan, Jiayang Gao, Andrew Gelman, Leslie Greengard,
Peter W. Jones, Deborah Mayo, Peter McCullagh, Michael O'Neil, Ron Peled,
William H. Press, Vladimir Rokhlin, Joseph Romano, Gary Simon, Amit Singer,
Michael Stein, Stephen Stigler, Joel Tropp, Larry Wasserman,
Douglas A. Wolfe, and Bin Yu.
This work was supported in part by Alfred P. Sloan Research Fellowships,
an NSF Postdoctoral Fellowship, a Donald D. Harrington Faculty Fellowship,
and a DARPA Young Faculty Award.

%%%\newpage

\addcontentsline{toc}{section}{\protect\numberline{}References}
\bibliographystyle{asamod.bst}
\bibliography{stat}

\begin{thebibliography}{12}
\newcommand{\enquote}[1]{``#1''}
\expandafter\ifx\csname natexlab\endcsname\relax\def\natexlab#1{#1}\fi

\bibitem[{Agresti(1992)}]{agresti1}
Agresti, A. (1992), A survey of exact inference for contingency tables,
  \textit{Statist. Sci.}, \textbf{7}, 131--153.

\bibitem[{Agresti(2001)}]{agresti2}
Agresti, A. (2001), Exact inference for categorical data: recent advances and
  continuing controversies, \textit{Stat. Med.}, \textbf{20}, 2709--2722.

\bibitem[{Bayarri and Berger(2004)}]{bayarri-berger2}
Bayarri, M.~J. and Berger, J.~O. (2004), The interplay of {B}ayesian and
  frequentist analysis, \textit{Statist. Sci.}, \textbf{19}, 58--80.

\bibitem[{Bickel et~al.(2006)Bickel, Ritov, and Stoker}]{bickel-ritov-stoker}
Bickel, P.~J., Ritov, Y., and Stoker, T.~M. (2006), Tailor-made tests for
  goodness of fit to semiparametric hypotheses, \textit{Ann. Statist.},
  \textbf{34}, 721--741.

\bibitem[{Clauset et~al.(2009)Clauset, Shalizi, and
  Newman}]{clauset-shalizi-newman}
Clauset, A., Shalizi, C.~R., and Newman, M. E.~J. (2009), Power-law
  distributions in empirical data, \textit{SIAM Review}, \textbf{51}, 661--703.

\bibitem[{Cox(2006)}]{cox}
Cox, D.~R. (2006), \textit{Principles of Statistical Inference}, Cambridge, UK:
  Cambridge University Press.

\bibitem[{Gelman(2003)}]{gelman}
Gelman, A. (2003), A {B}ayesian formulation of exploratory data analysis and
  goodness-of-fit testing, \textit{Internat. Stat. Rev.}, \textbf{71},
  369--382.

\bibitem[{Henze(1996)}]{henze}
Henze, N. (1996), Empirical-distribution-function goodness-of-fit tests for
  discrete models, \textit{Canad. J. Statist.}, \textbf{24}, 81--93.

\bibitem[{McCullagh and Nelder(1989)}]{mccullagh-nelder}
McCullagh, P. and Nelder, J.~A. (1989), \textit{Generalized Linear Models},
  London: Chapman and Hall, 2nd ed.

\bibitem[{Robins and Wasserman(2000)}]{robins-wasserman}
Robins, J.~M. and Wasserman, L. (2000), Conditioning, likelihood, and
  coherence: a review of some foundational concepts, \textit{J. Amer. Statist.
  Assoc.}, \textbf{95}, 1340--1346.

\bibitem[{Romano(1988)}]{romano}
Romano, J.~P. (1988), A bootstrap revival of some nonparametric distance tests,
  \textit{J. Amer. Statist. Assoc.}, \textbf{83}, 698--708.

\bibitem[{Ward(2012)}]{ward}
Ward, R.~A. (2012), Testing {H}ardy-{W}einberg equilibrium with a simple
  root-mean-square statistic, Tech. Rep. 12-24, UT-Austin Institute for
  Computational Engineering and Sciences, available at
  http://www.ices.utexas.edu/media/reports/2012/1224.pdf.

\end{thebibliography}

\end{document}